\documentclass[twocolumn,aps,showpacs,showkeys,preprintnumbers,]{revtex4}      
\usepackage{graphicx,url,graphics,epsfig,epsf,psfig}                                                        
\usepackage{dcolumn}                                                                                                             
\usepackage{amsmath}
\usepackage{amssymb}
\usepackage{bm}
\usepackage{longtable}                                                                                                                    
\usepackage[T1]{fontenc}

\begin{document}

%
%

\title{Valence-shell photoionization of chlorine-like Ar$^{+}$ ions}

\author{A. M. Covington}
\author{A. Aguilar}
\altaffiliation[Present address: ]{Advanced Light Source, Lawrence Berkeley National Laboratory,
             1 Cyclotron Road, MS 7-100, Berkeley, CA 94720, USA}
\author{I. R. Covington}
\author{G. Hinojosa}
\altaffiliation[Present address: ]{Instituto de Ciencias F\'{i}sicas, 
 						Universidad Nacional Aut\'onoma de M\'exico. 
						Apartado Postal 6-96, Cuernavaca 62131, Morelos, M\'exico.}

\author{C. A. Shirley}
\author{R. A. Phaneuf}
\affiliation{Department of Physics, MS 220, University of Nevada, Reno, NV 89557-0058, USA}

\author{I. \'{A}lvarez and C. Cisneros}
\affiliation{Instituto de Ciencias F\'{i}sicas, 
		Universidad Nacional Aut\'onoma de M\'exico,
		Apartado Postal 6-96, Cuernavaca 62131, Morelos, M\'exico.}

\author{I. Dominguez-Lopez}
 \altaffiliation[Present address: ]{Centro Nacional de Metrolog\'\i a,
                                                        Quer\'{e}taro, Qro. 76900, M\'{e}xico}
\author{M. M. Sant'Anna}
 \altaffiliation[Present address: ]{Instituto de F\'\i sica, Universidade Federal
               do Rio de Janeiro, Caixa Postal 68528, 21945-970 Rio de Janeiro RJ, Brazil}

\author{A. S. Schlachter}
\affiliation{Advanced Light Source, Lawrence Berkeley National Laboratory,
             1 Cyclotron Road, Berkeley, CA 94720, USA}

\author{C. P. Ballance}
\affiliation{Department of Physics, Auburn University, Auburn, AL 36840, USA}

\author{B. M. McLaughlin}
\altaffiliation[Present address: ]{Centre for Theoretical Atomic, Molecular and Optical Physics (CTAMOP), School of Mathematics and Physics,
                      The David Bates Building, 7 College Park, Queen's University of Belfast, Belfast BT7 1NN, United Kingdom}
\affiliation{Institute for Theoretical Atomic and Molecular Physics,
        Harvard Smithsonian Center for Astrophysics, 60 Garden Street, MS-14, Cambridge, MA 02138, USA}
%

\date{\today}

\begin{abstract}
Absolute cross-section measurements for valence-shell
photoionization of Ar$^{+}$ ions are reported for photon energies
ranging from 27.4 eV to 60.0 eV. The data, taken by merging beams
of ions and synchrotron radiation at a photon energy resolution of
10 meV, indicate that the primary ion beam was a statistically
weighted mixture of the $^2P^o_{3/2}$ ground state and the
$^2P^o_{1/2}$ metastable state of Ar$^{+}$. Photoionization of
this C$\ell$-like ion is characterized by multiple Rydberg series
of autoionizing resonances superimposed on a direct
photoionization continuum. Observed resonance lineshapes indicate
interference between indirect and direct photoionization channels.
Resonance features are spectroscopically assigned and their
energies and quantum defects are tabulated. The measurements are
satisfactorily reproduced by theoretical calculations based on an
intermediate coupling semi-relativistic Breit-Pauli approximation.
\end{abstract}


\maketitle
%
%
%
\section{Introduction}
Photoionization (PI) of ions is a fundamental process of
importance in many high-temperature plasma environments, such as
those occurring in stars and nebulae \cite{bregman} and in
inertial-confinement fusion experiments \cite{hofmann}.
Quantitative measurements of photoionization cross sections for
ions provide precision data on ionic structure and guidance to the
development of theoretical models of multi-electron interactions.
In addition, the opacity databases
\cite{topbase,opacity1,opacity2} that are critical to the modeling
and diagnostics of hot, dense plasmas consist almost entirely of
theoretical calculations performed in $LS$-coupling.
High-resolution absolute photoionization cross-section
measurements are therefore necessary to benchmark the validity of
this term-resolved data.

Due to the shortcomings of solar studies, abundance determinations
shed more light on the argon abundance at near-solar metallicity.
Such an approach has been suggested by recent studies on a similar
case of neon in B-type stars in the Orion nebula
\cite{cunha06,esteban04} and in the interstellar medium (ISM)
toward the Crab Nebulae \cite{kaastra07}. Photoionization
cross-section data on this C$\ell$-like ion are required to
accurately determine elemental abundances, for example in the
whole stellar sample of B-type stars where the $\lambda$~=~442.6
nm and $\lambda$~=~443.0 nm lines of Ar$^+$ are vividly seen in
the B main-sequence stars of the Orion nebulae \cite{lanz08}. The
non-LTE abundance calculations were based on photoionization cross
section data taken from the OPACITY PROJECT \cite{topbase}
calculated in $LS$-coupling. It has been shown for a variety of
ions, that the OPACITY PROJECT data has severe limitations when
compared to high resolution experimental data~
\cite{covington2002,aguilar2005}.

Photoionization phenomena are bound-free processes characterized
by interfering channels.  In direct photoionization of a positive
ion, the cross section rises from zero as a step-function at the
ionization threshold energy and falls off monotonically with
increasing photon energy.  This process  leads immediately to the
production of Ar$^{2+}$ and a free electron
\begin{equation}
\begin{array}{rcl}
        & (Ar^{+})^{**} & \\
   \nearrow &   & \searrow \\
\gamma + Ar^{+} & \longrightarrow & Ar^{2+} + e^{-}.
\label{eq:direct}
\end{array}
\end{equation}
Superimposed upon this direct photoionization cross section are
series of resonances occurring at discrete photon energies
corresponding to the excitation of autoionizing states
in the singly charged Ar$^+$ ion. Interference between the indirect and direct
photoionization channels produces characteristic Fano-Beutler line
profiles for the resonances \cite{fano}, providing further insight
into the electronic structure of the ion and the dynamics of the
photoionization process.

A thorough compilation of merged-beams photoionization studies has
been presented in a topical review by Kjeldsen~\cite{Kjeldsen2006}
and a previous review of West \cite{west}. The technique of
photoion-yield spectroscopy using synchrotron radiation was
pioneered by Lyon and collaborators \cite{lyon} at the Daresbury
synchrotron in the U.K., and involves merging ion beams
accelerated to keV energies with monochromatic, tunable beams of
synchrotron radiation.  Photoions produced by their interaction
over common paths of tens of centimeters may be magnetically or
electrostatically separated from the parent ion beam.  The
directed ion beam and negligible momentum transfer permit a
complete collection of the photoions. If the energy of the ion
beam is not too high and ultra-high vacuum conditions prevail, the
background produced by stripping collisions of the primary ion
beam with residual gas is manageable, and may be subtracted by
chopping the photon beam. The merged-beams technique is
particularly amenable to absolute cross-section measurements
\cite{phaneuf}.

Until recently, because of the relatively low density
presented by a space-charge-limited ion beam ($\sim$10$^{6}$ cm$^{-3}$),
data for ions were limited mainly to singly-charged alkali,
alkaline earth and transition metal ions with large
photoionization cross sections \cite{west}.
The advent of third-generation synchrotron radiation sources (3GLS)
with insertion devices increased the photon beam intensities
available to experimenters by nearly three orders of magnitude,
making possible the study of photoionization of ions with
unprecedented sensitivity and spectral resolution.
This has permitted measurements with ions of the lighter
and more astrophysically abundant elements, e.g.; C$^{+}$ \cite{kjeldsen},
O$^{+}$ \cite{covington2001,kjeldsen2}, C$^{2+}$ \cite{mueller2002} and more recently
C$^{3+}$ \cite{mueller2009}, N$^{3+}$ and O$^{4+}$ \cite{mueller2010b},
which are helping to refine theoretical descriptions of the
photoionization process.

As previously indicated the properties of Argon ions are of importance in atmospheric and
astrophysical plasmas and are widely used in
laboratory plasmas as diagnostic impurities \cite{plasma1, plasma2}.
Properties of Argon ions also find many applications in plasma
etching and lasers \cite{etching}.  In the present study
high-resolution absolute photoionization cross-section measurements and theoretical calculations
are presented for the Ar$^+$ ion at photon energies
ranging from the photoionization threshold to 60 eV.
%
%
%
%

\section{Experiment}

The experiment was conducted on undulator beamline 10.0.1.2 of the
Advanced Light Source (ALS) at Lawrence Berkeley National
Laboratory.  The ion-photon-beam (IPB) endstation is based on the
merged-beams technique and photoion spectroscopy using tunable
synchrotron radiation. The apparatus and methodology have been
described in detail previously for measurements of photoionization
cross sections for Ne$^+$ ions~\cite{covington2002}.
 Initial measurements of photoionization cross sections for
 O$^+$ ions \cite{covington2001} and C$^{2+}$ ions \cite{mueller2002}
using the IPB endstation were reported in the literature.
 The ALS group has recently reported detailed experimental and theoretical studies
 on the C$\ell$-like Calcium (Ca$^{3+}$) ion \cite{Phaneuf2010} in addition
 to a variety of other singly and multiply charged  ions,
 particularly of astrophysically interest, using this same IPB endstation;
He-like; Li$\rm ^{+}$ \cite{scully2006,scully2007},
Li-like; B$^{2+}$ \cite{mueller2010},
Be-like; C$\rm ^{2+}$  \cite{mueller2003},  B$\rm ^{+}$ \cite{schippers2003},
C$\rm ^{2+}$, N$\rm^{3+}$ and O$\rm^{4+}$ \cite{mueller2007,mueller2010b},
B-like; C$\rm ^{+}$  \cite{schlachter2004},
F-like; Ne$\rm ^{+}$  \cite{covington2002},
N-like ions; O$\rm ^{+}$ \cite{covington2001}, F$\rm ^{2+}$ and Ne$\rm ^{3+}$ \cite{aguilar2005}.
These high resolution experimental results enable theoretical studies to be benchmarked, so that the resulting
data can be used with confidence in various astrophysical applications that utilize
modeling codes such as Cloudy \cite{Ferland1998,Ferland2003} and XSTAR \cite{Kallman2001}.

\subsection{Photon beam}
The photon beam was produced by a 10-cm period undulator installed
in the 1.9 GeV electron storage ring to serve ALS Beamline 10.0.1.
A grazing-incidence spherical-grating monochromator delivered a
highly collimated photon beam of spatial width less than 1 mm and
divergence less than 0.5$^{\circ}$. The beamline produces a photon
flux of 5 $\times$ 10$^{12}$ photons per second in a bandwidth of
0.01$\%$ at an energy of 40 eV and three gratings cover the energy
range 17-340 eV.  Spectral resolving powers $E/\Delta E$ as high
as 40,000 are available at reduced photon flux. All of the
measurements with Ar$^{+}$ ions were carried out with a
gold-surfaced spherical grating ruled at 380 lines/mm. The photon
energy was scanned by rotating the grating and translating the
exit slit of the monochromator while simultaneously adjusting the
undulator gap to maximize the beam intensity. The spectral
resolution was pre-selected by adjusting the entrance and exit
slits of the monochromator.  The photon flux was measured by an
absolutely calibrated silicon X-ray photodiode, and was typically
2-3$\times$10$^{13}$ photons/second at a nominal spectral
resolving power of 2,000. The analog output from a precision
current meter was directed to a voltage-to-frequency converter,
which provided a normalization signal to a personal-computer-based
data acquisition system. The photon beam was time modulated
(mechanically chopped) at 0.5 Hz using a stepping-motor controlled
paddle to separate photoions from background produced by stripping
of the parent ion beam on residual gas in the ultra-high vacuum
system. The photon energy scale was calibrated using measurements
\cite{covington2001} of the well-known O$^{+}$ ground-state
($^{4}S$) and metastable-state ($^{2}P$ and $^{2}D$) energy
thresholds, allowing for the Doppler shift due to the ion motion
in the laboratory frame. The absolute uncertainty in the photon
energy scale is estimated to be $\pm$10 meV.

\subsection{Ion beam}
The $^{40}$Ar$^+$ ions were produced by the Cuernavaca Ion Gun
Apparatus (CIGA) in a hot-filament, low-pressure discharge-type
ion source, and accelerated to an energy of 6 keV.  The ion beam
was focused by a series of cylindrical electrostatic einzel lenses
and the Ar$^+$ ion beam selected by a 60$^{\circ}$ analyzing
magnet with a mass resolving power of 100.  The ion beam
trajectory and its cross-sectional area were defined by adjustable
horizontal and vertical beam slits located downstream of the
analyzing magnet. The collimated Ar$^+$ beam was typically a few
mm in diameter, with a current in the range 500-650 nA.  Three
stages of differential pumping with turbomolecular pumps and a
cryopump assured a downstream vacuum in the 10$^{-10}$ torr range
with beams present.

\subsection{Merger and beam interaction region}
A pair of 90$^{\circ}$ spherical-sector bending-plates merged the
ion beam onto the axis of the counter-propagating photon beam.
Fine-tuning of the overlap of the beams was achieved with two sets
of mutually perpendicular electrostatic steering plates mounted
immediately before the merger plates.  A cylindrical einzel lens
focused the beam in the center of the interaction region, which
consisted of an isolated stainless-steel-mesh cylinder to which an
electric potential (typically +2 kV) was applied, thereby
energy-labeling photoions produced in this region. Series of
entrance and exit apertures accurately defined the effective
length (29.4 cm) of the interaction region. Two-dimensional
intensity distributions of both beams were measured by commercial
rotating-wire beam profile monitors installed just upstream and
downstream of the interaction region, and by a translating-slit
scanner located in the middle of the region.  The profile monitors
permitted the positions and spatial profiles of the two beams to
be continuously monitored on an oscilloscope while tuning the
beams.  Two 500 l/s mag-ion pumps assured ultra-high vacuum
conditions in this region when the photon and ion beams were
present.

\subsection{Demerger and ion charge analyzer}
A 45$^{\circ}$ dipole analyzing magnet located downstream of the
interaction region demerged the beams and separated the Ar$^{2+}$
products from the parent Ar$^+$ beam, which was collected in an
extended Faraday cup. The magnetic field was set such that the
Ar$^{2+}$ product ions passed through an aperture in the back of
the Faraday cup.  A spherical 90$^{\circ}$ electrostatic deflector
directed them onto a stainless steel plate biased at -550V, from
which secondary electrons were accelerated and detected by a
microsphere-plate electron multiplier used in a pulse-counting
mode. The deflection planes of the demerger magnet and this
spherical deflector were orthogonal, permitting the Ar$^{2+}$
products to be swept across the detector in mutually perpendicular
directions, providing a diagnostic of their complete collection.
To this end, a cylindrical einzel lens located downstream of the
interaction region provided a further diagnostic, but was found to
be unnecessary and was turned off during the measurements.  The
absolute efficiency of the photoion detector was calibrated \it in
situ \rm using an averaging sub-femtoampere meter to record the
Ar$^{2+}$ photoion current, which was then compared to the
measured photoion count rate. The primary Ar$^{+}$ ion beam
current was measured by a precision current meter, whose analog
output was directed to a voltage-to-frequency converter, providing
a normalization signal to the data acquisition system.

%
%
%

\section{Theory}
For comparison with high-resolution measurements such as those at
ALS, state-of-the-art theoretical methods are required using
highly correlated wavefunctions. In addition, relativistic effects
are required since fine-structure effects can be resolved. As
metastable states are populated in the Ar$^+$ ion beam, additional
theoretical calculations were required to assist in the
determination of their fraction. Similar features have previously
been  demonstrated by past detailed experimental and theoretical
photoionization studies on a number of simple and complex ions, as
outlined in section II.

Previous photoionization cross-section calculations for Ar$^+$
were carried out in $LS$-coupling for the OPACITY PROJECT
\cite{topbase}. To provide a benchmark comparison and a guide to
the experiment, photoionization cross-section calculations were
performed in intermediate coupling (Breit-Pauli) using the
R-matrix method \cite{br75}. The inclusion of relativistic effects
is necessary due to the high energy resolution of the measurements
(nominally 10 meV). This theoretical approach optimizes use of
limited experimental beam time, since coarse energy scans may be
used in resonance-free regions and fine scans at high resolution
may be concentrated in energy ranges where dense resonance
structure is predicted. In the theoretical work twenty-four $LS$
terms of the product Ar$^{2+}$ ion were used, arising from the
configurations; $1s^22s^22p^63s^23p^4$, $1s^22s^22p^63s3p^5$,
 $1s^22s^22p^63s^23p^23d^2$ and $1s^22s^22p^63p^6$.
The high resolution spectral scan measurements show resonance
peaks with an energy separation of magnitude comparable to the
fine-structure splitting of the $1s^22s^22p^63s^23p^5~
^2P^o_{3/2,1/2}$ levels of Ar$^+$.

In the semi-relativistic R-matrix calculations performed for this complex, the 24 $LS$ terms
 give rise to 48 $LSJ$ levels for the product Ar$^{2+}$ ion core, all of which were included
in the close-coupling photoionization cross-section calculations.
For the Ar$^{2+}$ product ion the orbital basis set used to
describe the 24 target $LS$ terms
 was limited to n=3 in constructing the multi-configuration interaction wave functions used in the calculations.
Photoionization cross sections were  determined for both the
$^2P^o_{3/2}$ ground and the metastable $^2P^o_{1/2}$ initial
states of the Ar$^+$ ion due to the population of metastable
states in the primary ion beam used in the measurements.

An efficient parallel version \cite{ballance06} of the R-matrix
programs \cite{rmat,codes,damp} was used to determine all the photoionization cross
sections. The scattering wave functions were generated by allowing
for triple-electron promotions out of the base configuration
$3s^23p^5$ of Ar$^{+}$ into the n=3 orbital set
employed for the photoionization cross sections for both the
ground $^2P^o_{3/2}$ and the metastable $^2P^o_{1/2}$ initial
states of Ar$^{+}$ in intermediate coupling ($LSJ$).

All the scattering calculations were performed with twenty
continuum basis functions and a boundary radius of 7.0 Bohr radii.
In the case of the $^2P^o_{3/2}$ initial ground state, the dipole
selection rule requires the dipole transition matrices; $3/2^o
\rightarrow 1/2^e, 3/2^e, 5/2^e$, to be calculated, whereas for
the metastable $^2P^o_{1/2}$ inital state, only the dipole
matrices for the transitions; $1/2^o\rightarrow 1/2^e,3/2^e$ are
required. The Hamiltonian matrices for the $1/2^o$, $3/2^o$,
$5/2^e$, $3/2^e$ and $1/2^e$ symmetries were calculated over the
entire range of $LS$ partial waves that contribute to these $J\pi$
symmetries. The Ar$^{2+}$ theoretical energies were shifted to
experimental values in the diagonalization procedure of the
appropriate Hamiltonians so that the energy positions of
resonances relative to each ionization threshold would be
improved~\cite{NIST,hansen87}. This energy shift was less
than 1\% for all the energy levels involved. For both the
$^2P^o_{3/2}$ ground and the metastable $^2P^o_{1/2}$ initial
states, the electron-ion collision problem in the outer region was
then solved with the parallel R-matrix codes using a fine energy
mesh of 0.68 $\mu$eV (5 $\times$ 10$^{-8}$ Rydbergs). This ensured
that the fine resonance structure in the respective
photoionization cross sections was fully resolved, particularly in
the near-threshold energy region. The absolute theoretical
photoionization cross sections were convoluted with a Gaussian
with a FWHM corresponding to the experimental resolution
(nominally 10 meV) and statistically weighted in order to compare
directly with the experimental data.

The multi-channel R-matrix QB technique (applicable to atomic \cite{qb1,qb2} and molecular complexes \cite{qb3})
was used to determine the resonance parameters. In the QB method, the resonance width $\Gamma$
is determined from the inverse of the energy derivative of the eigenphase sum $\delta$
at the position of the resonance energy $E_r$ via
\begin{equation}
\Gamma = 2\left[{\frac{d\delta}{dE}}\right]^{-1}_{E=E_r} = 2 [\delta^{\prime}]^{-1}_{E=E_r} \quad.
\end{equation}
This approach exploits the analytical properties of the R-matrix
to obtain the energy derivative of the reactance ($\bf K$) matrix
and has previously proved very successful in locating fine
resonance features in the cross sections for multi-channel
electron scattering.

%
%
%
%

\section{Results}

\subsection{Absolute cross-section measurements}

Absolute measurements of photoionization cross sections were
performed at a number of discrete photon energies where there are
no resonant features in the photoion-yield spectrum. At each such
photon energy (\it h\rm $\nu$), the value of the total absolute
photoionization cross section $\sigma_{\rm PI}$ in cm$^2$ was
determined from experimentally-measured parameters:
\begin{equation}
\sigma_{\rm PI}(h\nu) = {\frac{R\,q\,{e^2}\,{v_i}\,
\epsilon}{I^+ I^\gamma \, \Omega \, \delta_1 \, \Delta \, {\int}F(z)dz}},
\label{eq:crossec}
\end{equation}
\noindent 
where $R$ is the photoion count rate [s$^{-1}$], $q$ is
the charge state of the parent ion, $e = 1.60 \times 10^{-19}$ C,
$v_{i}$ is the ion beam velocity [cm/s], $\epsilon$ is the
responsivity of the photodiode [electrons/photon], $I^{+}$ is the
ion beam current [A], $I^{\,\gamma}$ is the photodiode current
[A], $\Omega$ is the photoion collection efficiency, $\delta_1$ is
the pulse transmission fraction of the photoion detection
electronics (determined by the pulse-discriminator setting),
$\Delta$ is the measured absolute photoion detection efficiency,
and the beam overlap integral ${\int}F(z)dz$ defines the spatial
overlap of the photon and ion beams along the common interaction
path in units of cm$^{-1}$. The propagation direction of the ion
beam is defined as the z-axis. At each of the three positions
z$_i$ at which beam intensity profiles were measured, the form
factor $F(z_i)$ was determined by the following relation,

\begin{equation}
F(z_i) = {\frac{\int\int I^+(x,y)I^\gamma(x,y)dxdy}{\int\int I^+(x,y)dxdy \int\int I^\gamma(x,y)dxdy}}.
\label{eq:formfac}
\end{equation}
%
%
%
\begin{table}
\caption{\label{tab:param}Values of experimental parameters for a typical Ar$^{+}$ absolute 
 					photoionization cross-section measurement at 32.5 eV.}
\begin{ruledtabular}
\begin{tabular}{ll}
Parameter                                                           &Value(s)\\
[.02in]
\hline
Ion Beam Energy                                               &6.0 keV\\
Ion Beam Current, $I^+$                                  &620.0 nA\\
Photon Energy, $h{\nu}$                                  &32.4 eV\\
Photodiode current, $I^{\gamma}$                 &19.7 $\mu$A\\
Photon flux                                                          &5.6$\times$10$^{13}$ photons/s\\
Interaction bias voltage, $V_{int}$                  &+2.0 kV\\
Ion interaction velocity, $v_i$                          &1.39$\times$10$^7$ cm/s\\
Ar$^{2+}$ signal rate, $R$                              &7012.5 $s^{-1}$\\
Ar$^{2+}$ background rate                             &45.2 $s^{-1}$\\
Form Factors: $F(z_1)$, $F(z_2)$, $F(z_3)$        &41.22, 28.22, 19.17 cm$^{-2}$\\
Photodiode responsivity, $\epsilon$             &7.81 electrons/photon\\
Merge-path length, $L$                                   &29.4 cm\\
Pulse transmission fraction, $\delta_1$            &0.75\\
Photoion collection efficiency, $\Omega$   &1.00\\
Photoion detection efficiency, $\Delta$       &0.216\\
Cross Section, $\sigma$                                 &10.95 Mb \\
                                &(1Mb = $10^{-18}$ cm$^2$)\\
[0.01in]
\end{tabular}
\end{ruledtabular}
\end{table}

%
%

\begin{table}
\caption{\label{tab:abs}Systematic uncertainties in absolute
                                          cross-section measurements estimated at 90$\%$ confidence level.}
\begin{ruledtabular}
\begin{tabular}{lccc}
Source                                                    &Relative       &Absolute       &Total\\
\hline
Counting Statistics                                &2$\%$         &-                      &2$\%$\\
Photoion Detector Efficiency              &-                    &10$\%$            &10$\%$\\
Photoion Collection Efficiency           &2$\%$         &2$\%$            &3$\%$\\
Pulse Counting Efficiency                  &-                    &3$\%$            &3$\%$\\
Primary Ion Collection Efficiency      &-                    &2$\%$            &2$\%$\\
Ion Current Measurement                  &-                    &2$\%$            &2$\%$\\
Photodiode Responsivity                   &5$\%$         &10$\%$             &11$\%$\\
Photodiode Current Measurement   &2$\%$         &2$\%$            &3$\%$\\
Beam Profile Measurement               &3$\%$         &7$\%$            &8$\%$\\
Beam Overlap Integral                        &10$\%$      &7$\%$            &12$\%$\\
Interaction Length                               &-                    &2$\%$           &2$\%$\\
\hline
Quadrature Sum                                  &12$\%$      &18$\%$         &22$\%$\\
\end{tabular}
\end{ruledtabular}
\end{table}

\noindent 
The absolute photoion detection efficiency was determined in situ by comparing the 
count rate of Ar$^{2+}$ ions on the detector with the current measured using a 
calibrated averaging sub-femtoammeter. 
In Table  ~\ref{tab:param}, experimental parameters are given for a typical absolute cross-section 
measurement at a photon energy of 32.5 eV. 
Such measurements were repeated several times at each photon energy.

%
%

\begin{figure}
\begin{center}
\includegraphics[width=0.5\textwidth]{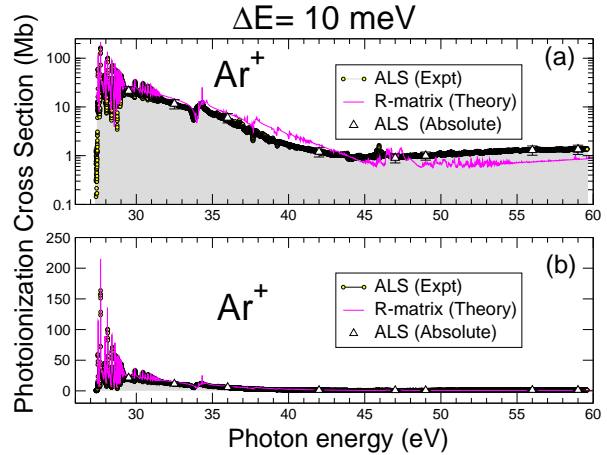}
\caption{\label{absolute} (Color online) Overview of experimental cross-section
                                           	measurements and theory for Ar$^+$ (shaded area) at a resolution
					of 10 meV over an energy range from the ionization threshold to 
					60 eV; (a) cross sections on a logarithmic scale, (b) cross section on a
					linear scale.  The absolute cross-section measurements are indicated by open
					triangles with error bars. The curve represents a 48-level
					close-coupling Breit-Pauli R-matrix calculation performed in
					intermediate coupling. For comparison, the theory curve is a
					statistically weighted sum of cross sections for photoionization
					from the ground and metastable states and has been convoluted with
					a FWHM Gaussian of 10 meV.}
\end{center}
\end{figure}
%
%
\begin{figure}
\begin{center}
\includegraphics[scale=0.325]{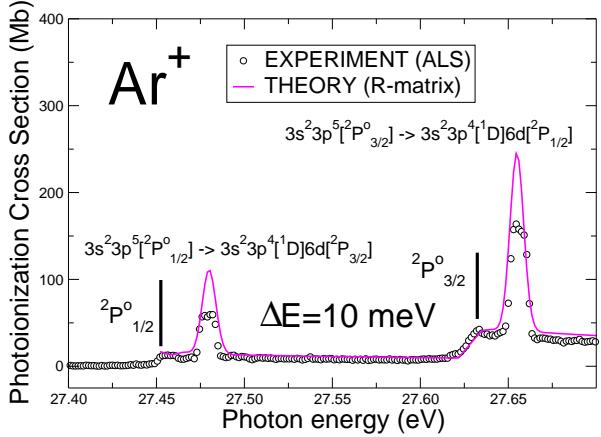}
\caption{\label{threshold} (Color online) High resolution PI cross-section measurements from the ALS at 10 meV
                                             energy resolution in the photon energy region of the $^2$P$_{3/2}$ ground-state
                                             and $^2$P$_{1/2}$ metastable-state ionization thresholds of Ar$^+$
                                             at 27.630 eV and 27.452 eV, respectively.  The small, non-zero cross
                                             section at energies below both thresholds is attributed to
                                             higher-order radiation in the photon beam.
                                             Theoretical results are from an intermediate coupling
                                             Breit-Pauli R-matrix calculation, statistically weighted for the ground and metastable state.
                                             The theoretical results have been convoluted with a Gaussian of 10 meV FWHM
                                             to simulate the photon energy bandwidth of the experiment.}
\end{center}
\end{figure}
Table~\ref{tab:abs} presents the relative, absolute and the total
uncertainties (their quadrature sums) in the photoionization
cross-section measurements estimated at 90$\%$ confidence level
(two standard deviations on statistical uncertainties) or at an
equivalent confidence level for systematic uncertainties.
Table~\ref{tab:table3} presents the absolute photoionization
cross-section values that were determined at the selected photon
energies. The data from Table~\ref{tab:table3} are plotted in
Fig.~\ref{absolute} with error bars representing the total
uncertainties. Photon energy scans taken over the energy range
27.4 -- 60 eV at a resolution of 10 meV with 1.5 meV steps were
normalized to these absolute measurements by fitting a polynomial
to their measured ratio as a function of photon energy. These data
are presented in Fig.~\ref{absolute} as small triangles.

%
%
\begin{figure}
\begin{center}
\includegraphics[scale=0.45]{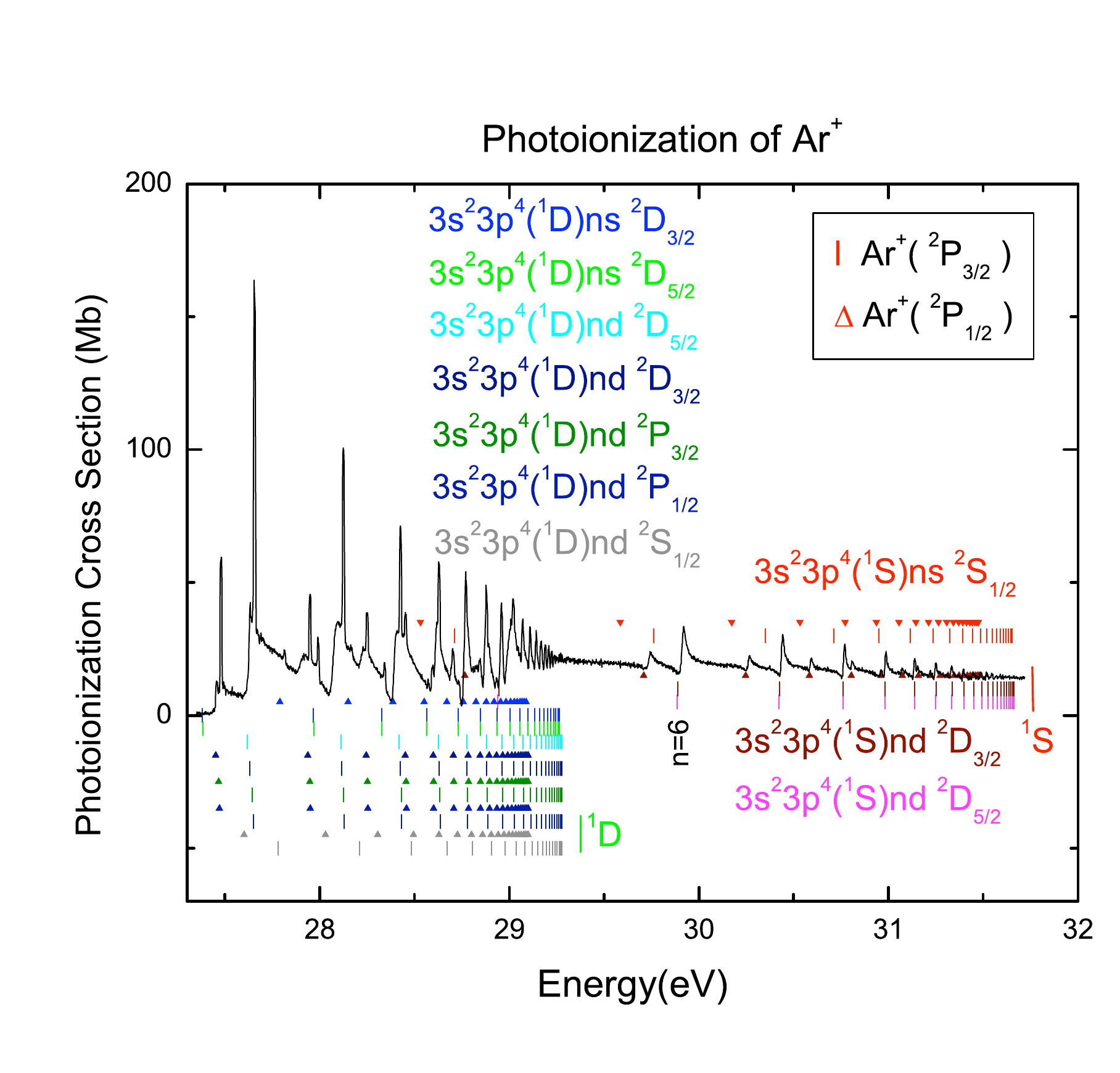}
\caption{\label{expt2} (Color online) Experimental results from the ALS for the PI cross sections of
                                           Ar$^+$ for photon energies ranging from 27-32 eV
                                           with a photon energy resolution of 10 meV. In the figure,  the
                                           members of Rydberg series of resonances
                                           originating from the $^{2}P^o_{1/2}$ metastable ($\mid$) and
                                           $^{2}P^o_{3/2}$ ground states ($\bigtriangleup$) of Ar$^+$ are
                                           identified. The resonance assignments were determined from the NIST tabulated levels
                                           for Ar II and Ar III \cite{NIST}.}
\end{center}
\end{figure}

%
%
%
\begin{figure}
\begin{center}
\includegraphics[scale=0.3545]{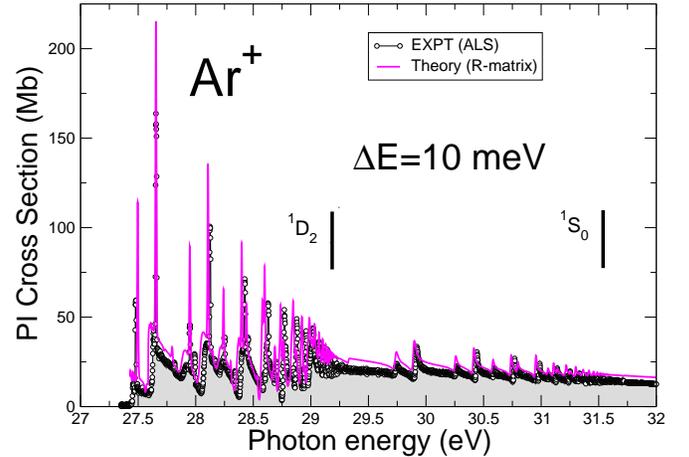}
\caption{\label{expt3} (Color online) Experimental results from the ALS for the PI cross sections of
                                        Ar$^+$ for photon energies ranging from 27-32 eV
                                       with a photon energy resolution of 10 meV compared
                                       with theoretical results from a 48-state Breit-Pauli R-matrix calculation.
                                       The theoretical results have been convoluted with a  Gaussian of FWHW of 10 meV and
                                       statistically averaged over the ground and metastable states.}
\end{center}
\end{figure}

%

\begin{table}
\caption{\label{tab:table3}Measured values of the total
                                              absolute cross sections for photoionization of Ar$^+$.}
\begin{ruledtabular}
\begin{tabular}{ccc}
Energy (eV)          &Cross Section                             &Total Uncertainty\\
(eV)                        &(10$^{-18}$ cm$^2$)                &(10$^{-18}$cm$^2$)\\
\hline\\
29.5                       &21.4                                              &4.6\\
32.5                       &11.6                                              &2.5\\
36.0                       &~6.0                                              &1.3\\
42.0                       &~1.2                                              &0.3\\
47.0                       &~0.9                                              &0.2\\
49.0                       &~1.0                                              &0.2\\
56.0                       &~1.3                                              &0.3\\
59.0                       &~1.4                                              &0.3\\
\end{tabular}
\end{ruledtabular}
\end{table}


%
%
\begin{figure}
\begin{center}
\includegraphics[scale=0.355]{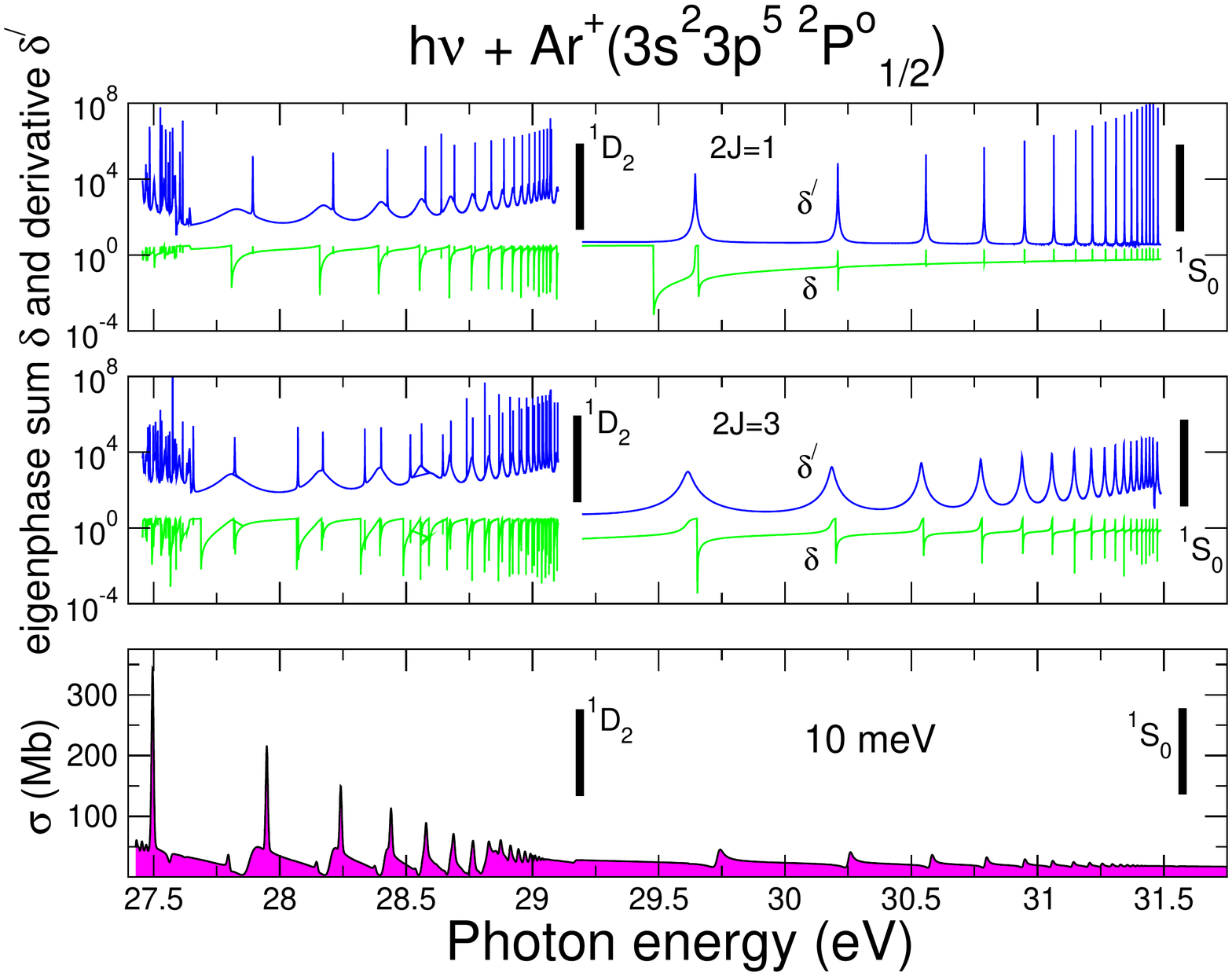}
\caption{\label{phase1} (Color online)
				      Eigenphase sum $\delta$ and its derivative $\delta^{\prime}$ for
				      each $2J^{\pi}$ contributing to
                                            the total PI cross section $\sigma$  originating
                                            from the initial Ar$^{+}$($^2P^o_{1/2}$)  metastable state.
                                            The PI cross section has been convoluted with a FWHM Gaussian 
                                            of 10 meV to simulate the photon experimental bandwidth.}
\end{center}
\end{figure}

%
%
\begin{figure}
\begin{center}
\includegraphics[scale=0.355]{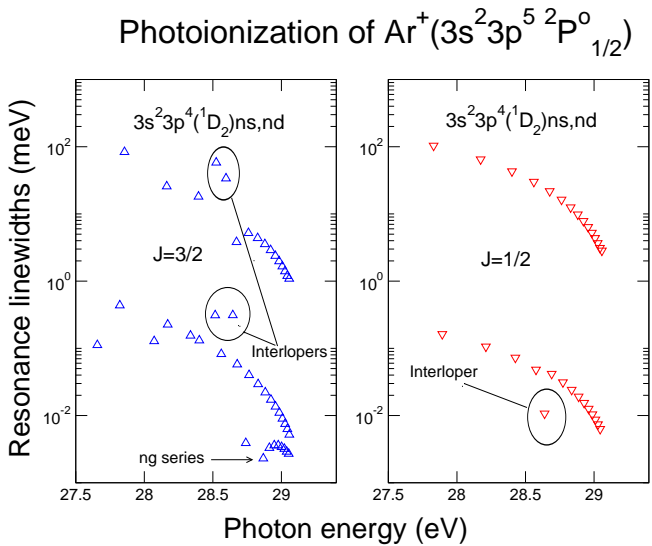}
\caption{\label{2p1-reson1} (Color online) Linewidths (meV) for the dominant $\:3s^23p^4(^1D_2)ns,nd$ resonances series
                                                   converging to the $^1D_2$ threshold of the product Ar$^{2+}$ ion originating
                                                   from the initial Ar$^{+}$($^2P^o_{1/2}$) metastable state. 
                                                   The $\:3s^23p^4(^1D_2)ng$ Rydbergy resonance series have 
                                                   linewidths very much less than 10$^{-2}$ meV and have not been analyzed.}
\end{center}
\end{figure}

%
 \begin{figure}
\begin{center}
\includegraphics[scale=0.355]{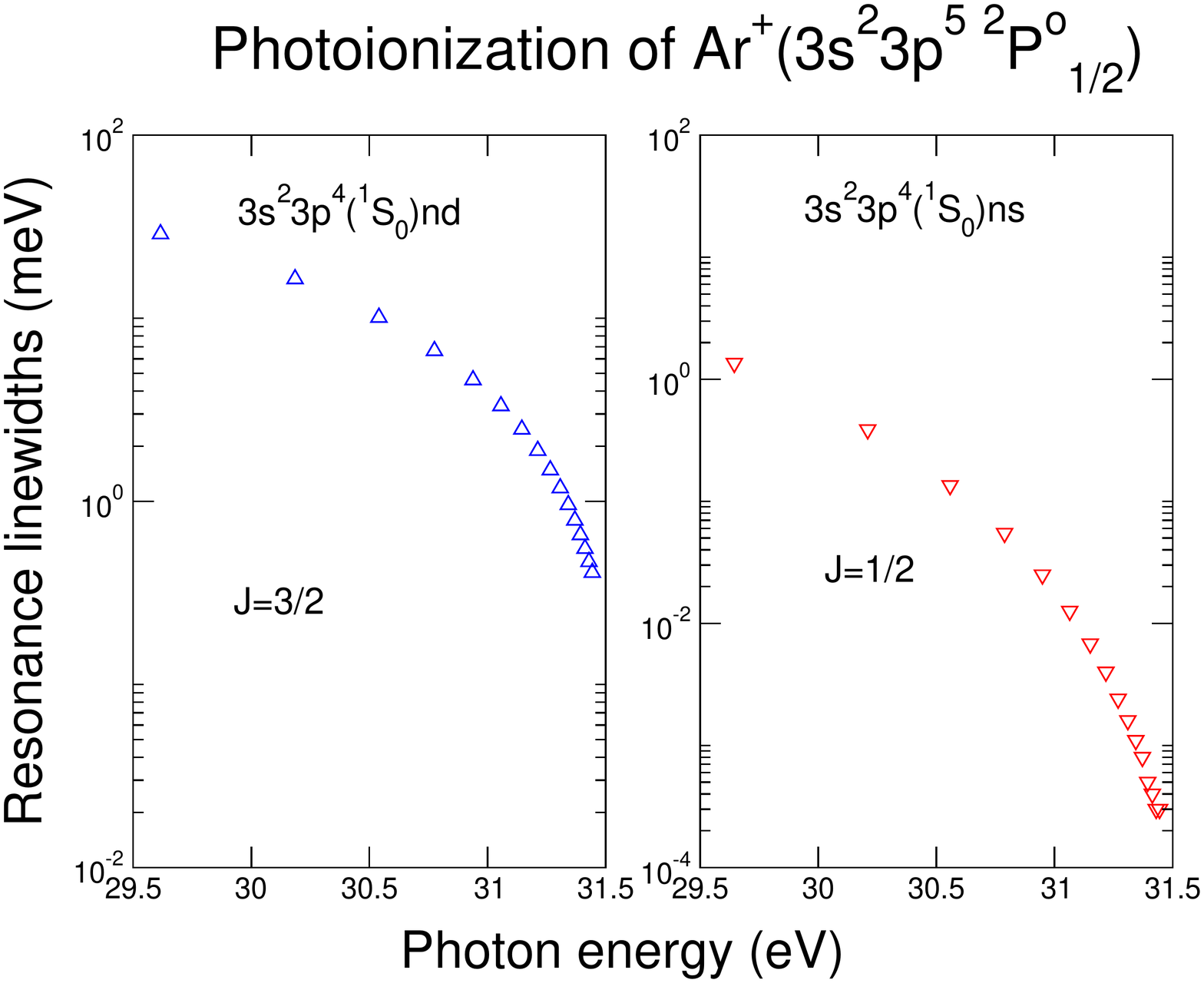}
\caption{\label{2p1-reson2}  (Color online)  Linewidths (meV) for the  dominant $\:3s^23p^4(^1S_0)ns,nd$ resonances series
                                                     converging to the $^1S_0$ threshold of the product Ar$^{2+}$ ion originating
                                                     from the initial Ar$^{+}$($^2P^o_{1/2}$) metastable state.}
\end{center}
\end{figure}
%
%
\begin{figure}
\begin{center}
\includegraphics[scale=0.355]{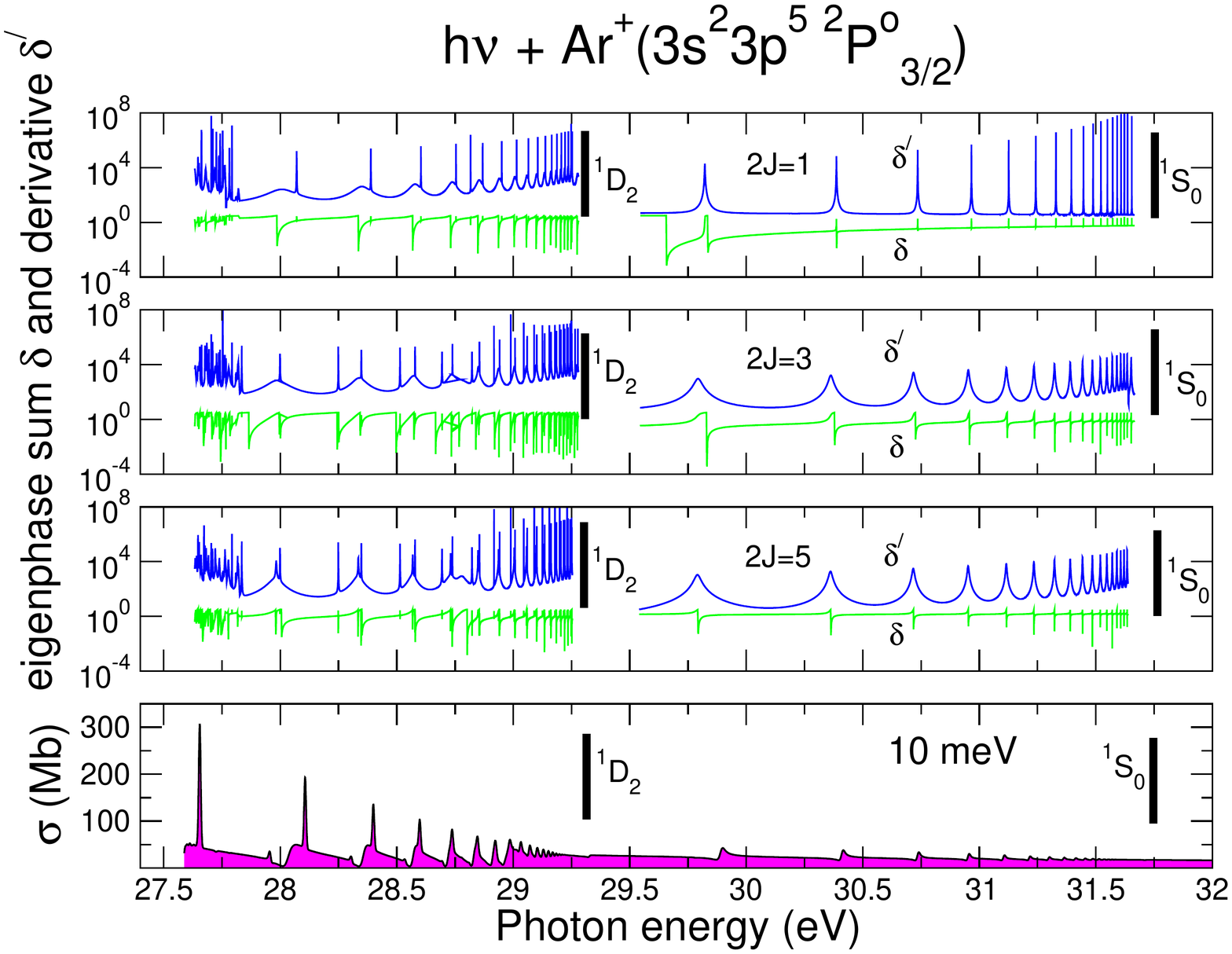}
\caption{\label{phase2} (Color online) Eigenphase sum $\delta$ and its derivative $\delta^{\prime}$ for each $2J^{\pi}$ contributing to
                                            the total  PI cross section $\sigma$  originating
                                            from the  initial Ar$^{+}$($^2P^o_{3/2}$) ground state.
                                            The PI cross section has been convoluted with a FWHM 
                                            Gaussian of 10 meV to simulate the photon experimental bandwidth.}
\end{center}
\end{figure}
%
%
 \begin{figure}
\begin{center}
\includegraphics[scale=0.355]{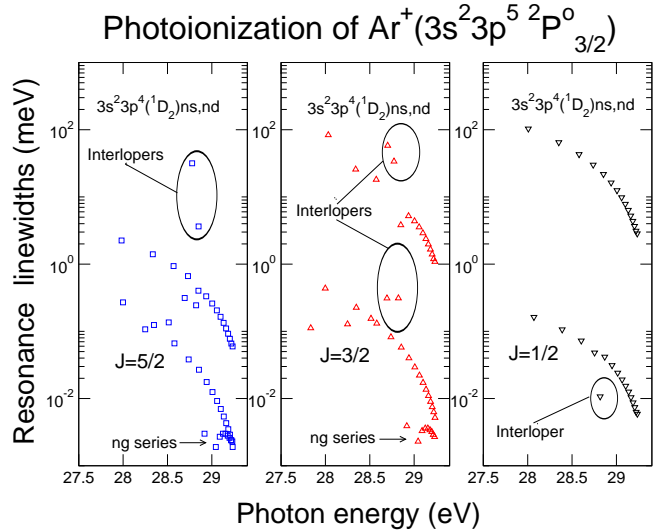}
\caption{\label{2p3-reson1} (Color online) Linewidths (meV) for the dominant $\:3s^23p^4(^1D_2)ns,nd$ resonances series converging
                                                   to the $^1D_2$ threshold of the product Ar$^{2+}$ ion originating
                                                   from photoionization of the initial Ar$^{+}$($^2P^o_{3/2}$) ground state.
                                                   The $\:3s^23p^4(^1D_2)ng$ Rydbergy resonance series have 
                                                   linewidths very much less than 10$^{-2}$ meV and have not been analyzed.}
\end{center}
\end{figure}

%
%
 
 \begin{figure}
\begin{center}
\includegraphics[scale=0.355]{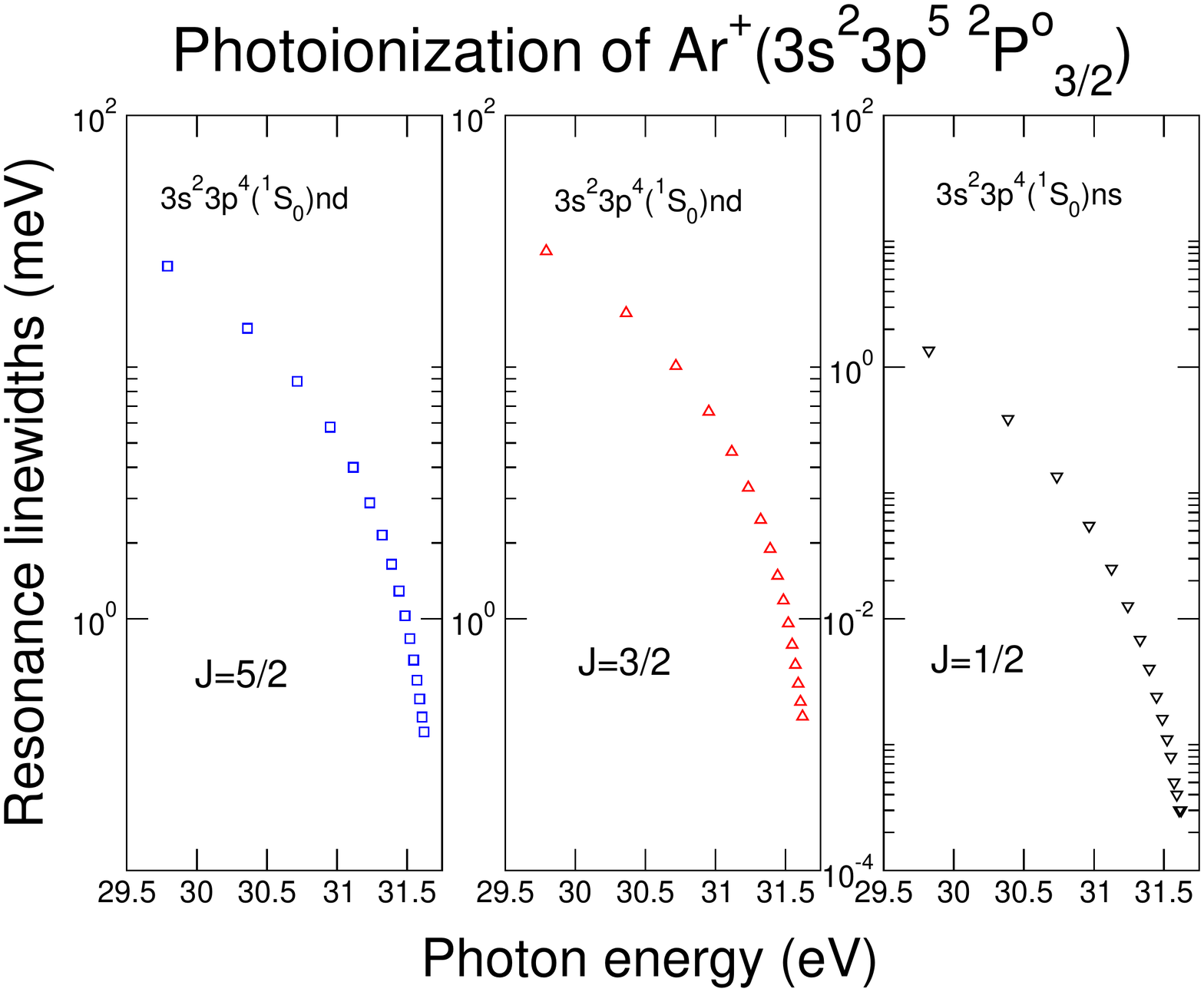}
\caption{\label{2p3-reson2}  (Color online) Linewidths (meV) for the dominant $\:3s^23p^4(^1S_0)ns,nd$ resonances series converging
                                                   to the $^1S_0$ threshold of the product Ar$^{2+}$ ion originating
                                                   from photoionization of the  initial Ar$^{+}$($^2P^o_{3/2}$) ground state.}
\end{center}
\end{figure}

Fig.~\ref{threshold} presents the ALS measured photoionization
cross section in the region of the $^2$P$_{3/2}$ ground-state and
$^2$P$_{1/2}$ metastable-state ionization threshold energies at
27.630 eV and 27.452 eV, respectively according to those given by
the NIST tabulations \cite{NIST}. The ratio of the magnitudes of
the two threshold cross-section steps is 2.06 $\pm$ 0.10,
	consistent with a statistical population (proportional to $2J + 1$) of
ground- and metastable-state ions in the primary ion beam.  The
present measurements are therefore considered to correspond to a
sum of the $^2$P$_{3/2}$ ground-state cross section multiplied by
2/3 and the metastable-state $^2$P$_{1/2}$ cross section
multiplied by 1/3. The presence of the metastable component adds
complexity to the observed resonance structure and to its
interpretation.  A very small offset of the measured cross section
from zero below the $^2$P$_{1/2}$ threshold is attributed to the
presence of higher-order radiation from the undulator, which is
dispersed by the grating and estimated to comprise 6$\%$ of the
total photon flux in this energy region. In this threshold region,
a detailed comparison of the ALS experimental results with
theoretical results (from a Breit-Pauli calculation performed in
intermediate coupling) show excellent agreement. In order to make
a direct comparison with experiment in this energy region the
theoretical results  were convoluted with a FWHM gaussian of 10
meV and statistically weighted over the ground and metastable
states. The two strong resonance features observed in the
threshold energy region of the ALS photoionization experimental
measurements, (located at 27.481 eV and 27.658 eV 
respectively; see Fig~\ref{threshold} )
resulting in photoexcitation of the $\:3s^23p^4(^1D_2)6d$ Auger
state, (from either the ground state or metastable states of Ar$^{+}$),
are reproduced suitably well by the present theoretical
Breit-Pauli calculations performed in intermediate coupling. 

The peak heights of the stronger calculated resonances exceed the measured values, 
and the experimental lineshapes of the resonances at 27.48 eV and 27.66 eV are non-Gaussian. 
This suggests a possible saturation effect due to the dead time 
of the photoion counting system at the higher signal counting rates. 
The measured ratios of the strengths of these two resonance further 
suggests that the ground-state fraction in the primary ion beam may 
have been closer to 0.73 than the value of 0.67 based on statistical 
weights that was used to scale the theoretical results for comparison with experiment.

In Fig.~\ref{expt2} we present  the
experimental data taken at 10 meV in the photon energy region from threshold to approximately 32 eV.
As can been seen from Fig.~\ref{expt2}, there is extremely rich resonance structure evident in photoionization cross section data.
The assignments of the various Rydberg resonance series are given based on the  NIST tabulated levels for Ar II and Ar III \cite{NIST}.
Only the prominent resonance series are analyzed in detail.  Fig.~\ref{expt3} illustrates the same energy
range but includes the comparison with the theoretical results obtained  from the intermediate coupling Breit-Pauli calculations.
Here again theory has been convoluted with a FWHM Gaussian of 10 meV and statistically weighted
over the ground and metastable states to make a direct comparison with experiment in this energy region.
Once again it is seen there is remarkable agreement between theory and experiment,
both on the absolute cross section scale and on the photon energy scale.

\subsection{Resonances}

In the photon energy range 27 eV to  32 eV a wealth of resonance
structure is observed in the PI cross sections as illustrated in
Fig. 3.  We note that below the $^1D_2$ threshold of the product
Ar$^{2+}$ ion, $3s^23p^4(^1D_2)ns~~^2D_{3/2,5/2}$,
$3s^23p^4(^1D_2)nd~^2D_{3/2,5/2}$,
$3s^23p^4(^1D_2)nd~^2P_{1/2,3/2}$ and
$3s^23p^4(^1D_2)nd~^2S_{1/2}$ dominant Rydberg resonance series may occur
from the angular momentum coupling rules.

At higher photon energies $3s^23p^4(^1S_0)nd~^2D_{3/2,5/2}$ and
$3s^23p^4(^1S_0)ns~^2S_{1/2}$ Rydberg resonance series converging
to the $^1S_0$ threshold of the corresponding product Ar$^{2+}$
ion are observed.  The resonance series found are illustrated in
Figs. 5-10.  For resonances converging to both the $^1D_2$  and
$^1S_0$ thresholds of the Ar$^{2+}$ ion only the dominant resonances
series are analyzed.

To assist in the identification of the dominant resonance features observed 
in the photoionization spectra a theoretical analysis was carried
out based on the widely used QB technique for determining
resonance parameters in atomic and molecular systems
\cite{qb1,qb2,qb3}. This technique exploits the properties of the
R-matrix in multi-channel scattering and is particularly helpful
for locating and determining the properties of narrow resonances found in the spectra.

The relationship between the principal quantum number n,
the effective quantum number $\nu$ and the quantum defect $\mu$
for an ion of effective charge ${\cal Z}$ is given by
$\nu$ = n - $\mu$ where the resonance position $\epsilon_r$ can be determined from Rydberg's formula
\begin{equation}
\epsilon_r  =  \epsilon_{\infty} -  {\cal~Z}^2 / \nu^{2},
\end{equation}
 where $\epsilon_{\infty}$ is the resonance series limit.

Prominent Rydberg resonance series observed in the experimental
photoionization spectra have been analyzed and assigned according
to their quantum defects. The NIST tabulations were used as a
helpful guide in the assignment. The principal quantum number n,
resonance energy in eV, quantum defect $\mu$ and the resonance
autoionizing linewidth $\Gamma$ (meV) of the first few members of
the dominant Rydberg series originating from the $\rm ^2P^o_{1/2}$
metastable and the $\rm ^2P^o_{3/2}$  ground state of the Ar$^{+}$
ion are listed in the Tables IV, V and VI. The results for the
appropriate resonance parameters for the dominant Rydberg
resonance series found in the spectra of the PI cross section for
the ground and metastable states of Ar$^{+}$, lying below the
$^1D_2$  and $^1S_0$  thresholds of the product Ar$^{2+}$ ion are
listed in Tables IV, V and VI. As may be seen from the results of
the QB analysis, many members of the various Rydberg series have
resonance linewidths smaller than the experimental photon
bandwidth of 10 meV, therefore making their detection extremely
difficult. Finally we note that interloping resonances 
$\:3s^23p^4(^1S_0)6s,5d$ are present in the spectrum below 
the  $\:3s^23p^4(^1D_2)$ threshold (as clearly  illustrated  in figures 6 and 9) 
disrupting the regular pattern of the Rydberg series.

%
%
%

\begin{table*}
\caption{\label{tab:resonances1}Principal quantum numbers $n$, resonance energies (eV), quantum defects
                                                         $\mu$ and autoionization linewidths [$\Gamma$(meV)], determined by the QB method for the  dominant
                                                         $3s^23p^4(^1S_0)ns,nd$ series observed in the photoionization spectra originating from the $3s^23p^5\,^2P^o_{1/2}$
                                                         metastable state of Ar$^{+}$.}
\begin{ruledtabular}
\begin{tabular}{cccccccccc}
                    &&      &Rydberg Series         &       &Width ($\Gamma$)   &Rydberg Series     &       & Width ($\Gamma$)\\[.02in]
                    & &     &$\:3s^23p^4(^1D_2)ns,nd$   &       &$\:3s^23p^4(^1D_2)ns,nd$&$\:3s^23p^4(^1S_0)ns,nd$&& $\:3s^23p^4(^1S_0)ns,nd$\\
Initial State           & &$n$  &Energy (eV)                &$\mu$  & Energy(meV)           &Energy (eV)            &$\mu$  & Energy (meV)  \\[.02in]
\hline\\
                    &&      &                       &j=1/2  &                   &                        &      &           \\
$3s^23p^5\,^2P^o_{1/2}$ &&[7s]  &-                      &-      &-                  &29.645             &1.692  &~1.36      \\
                    &&8     &27.830                 &1.673  &103.4              &30.209             &1.691  &~0.39      \\
                    &&9     &28.173                 &1.682  &~64.8              &30.558             &1.690  &~0.14      \\
                    &&10    &28.401                 &1.688  &~43.0              &30.789             &1.689  &~0.06      \\
                    &&11    &28.561                 &1.692  &~30.0              &30.949             &1.688  &~0.03      \\
                    &&12    &28.677                 &1.695  &~21.8              &31.065             &1.688  &~0.01      \\
                    &&13    &28.763                 &1.698  &~16.3              &31.151             &1.688  &< 10$^{-2}$    \\
                    &&14    &28.829                 &1.699  &~12.5              &31.217             &1.688  &< 10$^{-2}$    \\
                    &&15    &28.881                 &1.701  &~~9.8              &31.269             &1.688  &< 10$^{-2}$    \\
                    &&16    &28.923                 &1.702  &~~7.8              &31.311             &1.687  &< 10$^{-2}$    \\
                    &&$\cdot$ &$\cdot$                  &$\cdot$    &                   &$\cdot$                &$\cdot$    & -         \\
                    &&$\infty$  &29.189                 &-      &                   &31.596             &-      & -         \\
                    \\
                    &&      &                       &j=1/2  &                   &                        &      &           \\
$3s^23p^5\,^2P^o_{1/2}$&&[6d]   &-                      &-      &-                  &                   &       &           \\
                    &&7     &27.830                 &0.521  &0.16               &                   &       &           \\
                    &&8     &28.211                 &0.540  &0.11               &                   &       &           \\
                    &&9     &28.426                 &0.554  &0.07               &                   &       &           \\
                    &&10    &28.576                 &0.574  &0.05               &                   &       &           \\
                    &&11    &28.691                 &0.546  &0.04               &                   &       &           \\
                    &&12    &28.773                 &0.562  &0.03               &                   &       &           \\
                    &&13    &28.837                 &0.568  &0.02               &                   &       &           \\
                    &&14    &28.887                 &0.572  &0.02               &                   &       &           \\
                    &&15    &28.928                 &0.574  &0.02               &                   &       &           \\
                    &&$\cdot$ &$\cdot$                  &$\cdot$    &                   &                   &       &           \\
                    &&$\infty$  &29.189                 &-      &                   &                   &       &           \\
                    \\
                    &&      &                       &j = 3/2    &                   &                   &       &           \\
$3s^23p^5\,^2P^o_{1/2}$ &&[7s]  &-                      &-      &                   &                   &       &           \\
                    &&8     &27.821                 &1.692  &0.44               &                   &       &           \\
                    &&9     &28.171                 &1.691  &0.22               &                   &       &           \\
                    &&10    &28.401                 &1.690  &0.13               &                   &       &           \\
                    &&11    &28.561                 &1.689  &0.08               &                   &       &           \\
                    &&12    &28.677                 &1.688  &0.08               &                   &       &           \\
                    &&13    &28.764                 &1.688  &0.04               &                   &       &           \\
                    &&14    &28.830                 &1.688  &0.02               &                   &       &           \\
                    &&15    &28.882                 &1.687  &0.02               &                   &       &           \\
                    &&16    &28.923                 &1.687  &0.02               &                   &       &           \\
                    &$\cdot$    &$\cdot$                    &$\cdot$    &                   &                   &       &           \\
                    &&$\infty$   &29.189                    &-      &                   &                   &       &           \\
                    \\
                    &&      &                       &j = 3/2    &                   &                   &       &           \\
$3s^23p^5\,^2P^o_{1/2}$ &&[6d]  &-                      &-      &                   &29.615             &0.732  &28.9       \\
                    &&7     &27.855                 &0.613  &~83.4              &30.185             &0.746  &16.4       \\
                    &&8     &28.163                 &0.716  &~25.7              &30.540             &0.754  &10.1       \\
                    &&9     &28.395                 &0.720  &~18.1              &30.775             &0.759  &~6.6       \\
                    &&10    &(28.590)              &(0.422)    &~33.8              &30.938             &0.763  &~4.6       \\
                    &&11    &28.672                 &0.738  &~~3.8              &31.057             &0.766  &~3.3       \\
                    &&12    &28.760                 &0.745  &~~5.3              &31.145             &0.767  &~2.5       \\
                    &&13    &28.827                 &0.744  &~~4.4              &31.213             &0.769  &~1.9       \\
                    &&14    &28.879                 &0.744  &~~3.6              &31.266             &0.770  &~1.5       \\
                    &&15    &28.921                 &0.744  &~~2.9              &31.308             &0.771  &~1.2       \\
                    &&$\cdot$ &$\cdot$                  &$\cdot$    &                   &$\cdot$                &$\cdot$    &-          \\
&&$\infty$\footnote{Series limit from NIST Standard Reference Database \protect\cite{NIST}.}&29.189     &-  &       &31.596 &-          &           \\
\end{tabular}
\end{ruledtabular}
\end{table*}

%
%

\begin{table*}
\caption{\label{tab:resonances2}Principal quantum numbers $n$, resonance energies (eV), quantum defects
                                                         $\mu$ and autoionization linewidths [$\Gamma$(meV)], determined from the QB method for
                                                         the dominant $\:3s^23p^4(^1D_2)ns,nd$ and $3s^23p^4(^1S_0)ns,nd$ series resulting from
                                                         photoionization of the $3s^23p^5\,^2P^o_{3/2}$ ground state of Ar$^{+}$.}
\begin{ruledtabular}
\begin{tabular}{cccccccccc}
                    &&      &Rydberg Series         &       &Width ($\Gamma$)   &Rydberg Series     &       & Width ($\Gamma$)\\[.02in]
                    &&      &$\:3s^23p^4(^1D_2)ns,nd$   &       &$\:3s^23p^4(^1D_2)ns,nd$   &$\:3s^23p^4(^1S_0)ns,nd$&  & $\:3s^23p^4(^1S_0)ns,nd$\\
Initial State           &&$n$   &Energy (eV)                &$\mu$  & Energy(meV)       &Energy (eV)            &$\mu$  & Energy (meV)  \\[.02in]
\hline\\
                    &&      &                       &j = 1/2    &               &                   &       &            \\
$3s^23p^5\,^2P^o_{3/2}$ &&[7s]  &-                      &-      & -             &29.822             &1.692  &~1.36      \\
                    &&8     &28.007                 &1.673  &103.4          &30.387             &1.691  &~0.39      \\
                    &&9     &28.351                 &1.682  &~64.8          &30.736             &1.690  &~0.14      \\
                    &&10    &28.579                 &1.688  &~43.0          &30.966             &1.689  &~0.06      \\
                    &&11    &28.739                 &1.692  &~30.0          &31.126             &1.689  &~0.03      \\
                    &&12    &28.854                 &1.696  &~21.8          &31.242             &1.688  &~0.01      \\
                    &&13    &28.941                 &1.698  &~16.3          &31.329             &1.688  &< 10$^{-2}$    \\
                    &&14    &29.007                 &1.700  &~12.5          &31.395             &1.688  &< 10$^{-2}$    \\
                    &&15    &29.059                 &1.701  &~~9.8          &31.447             &1.688  &< 10$^{-2}$    \\
                    &&16    &29.100                 &1.702  &~~7.8          &31.488             &1.688  &< 10$^{-2}$    \\
                    &&$\cdot$ &$\cdot$                  &$\cdot$    &               &$\cdot$                &$\cdot$    &-          \\
                    &&$\infty$  &29.367                 &-      &               &31.774             &-      &-          \\
                    \\
                    &&      &                       &j=1/2  &               &                        &      &           \\
$3s^23p^5\,^2P^o_{3/2}$&&[6d]   &-                      &-      &-              &                   &       &           \\
                    &&7     &28.070                 &0.521  &0.16           &                   &       &           \\
                    &&8     &28.389                 &0.540  &0.11           &                   &       &           \\
                    &&9     &28.603                 &0.554  &0.07           &                   &       &           \\
                    &&10    &28.754                 &0.574  &0.05           &                   &       &           \\
                    &&11    &28.869                 &0.546  &0.04           &                   &       &           \\
                    &&12    &28.951                 &0.562  &0.03           &                   &       &           \\
                    &&13    &29.014                 &0.568  &0.02           &                   &       &           \\
                    &&14    &29.065                 &0.572  &0.02           &                   &       &           \\
                    &&15    &29.105                 &0.574  &0.02           &                   &       &           \\
                    &&$\cdot$ &$\cdot$                  &$\cdot$    &               &                   &       &           \\
                    &&$\infty$  &29.367                 &-      &               &                   &       &           \\
                    \\
                    &&      &                       &j=3/2  &               &                   &       &           \\
$3s^23p^5\,^2P^o_{3/2}$&&[7s]   &-                      &-      &-              &                   &       &           \\
                    &&8     &27.999                 &1.692  &0.44           &                   &       &           \\
                    &&9     &28.348                 &1.691  &0.26           &                   &       &           \\
                    &&10    &28.579                 &1.690  &0.13           &                   &       &           \\
                    &&11    &28.739                 &1.689  &0.08           &                   &       &           \\
                    &&12    &28.855                 &1.688  &0.06           &                   &       &           \\
                    &&13    &28.941                 &1.688  &0.04           &                   &       &           \\
                    &&14    &29.098                 &1.688  &0.03           &                   &       &           \\
                    &&15    &29.060                 &1.687  &0.02           &                   &       &           \\
                    &&16    &29.101                 &1.687  &0.02           &                   &       &           \\
                    &&$\cdot$ &$\cdot$                  &$\cdot$    &               &                   &       &           \\
                    &&$\infty$  &29.367                 &-      &               &                   &       &           \\
                    \\
                    &&      &                       &j=3/2  &               &                   &       &           \\
$3s^23p^5\,^2P^o_{3/2}$&&[6d]   &-                      &-      &-              &29.793             &0.732  &28.9       \\
                    &&7     &28.033                 &0.613  &~83.4          &30.363             &0.746  &16.4       \\
                    &&8     &28.341                 &0.716  &~25.7          &30.718             &0.754  &10.1       \\
                    &&9         &28.573                 &0.720  &~18.1          &30.953             &0.759  &~6.6       \\
                    &&10    &(28.774)                   &(0.422)    &~33.8          &31.116             &0.763  &~4.6       \\
                    &&11    &28.850                 &0.738  &~~3.8          &31.234             &0.766  &~3.3       \\
                    &&12    &28.937                 &0.745  &~~5.3          &31.322             &0.768  &~2.5       \\
                    &&13    &29.004                 &0.744  &~~4.4          &31.390             &0.769  &~1.9       \\
                    &&14    &29.047                 &0.744  &~~3.6          &31.443             &0.770  &~1.5       \\
                    &&15    &29.099                 &0.744  &~~2.9          &31.485             &0.771  &~1.2       \\
                    &&$\cdot$   &$\cdot$                    &$\cdot$    &-              &$\cdot$                &$\cdot$    &           \\
&&$\infty$\footnote{Series limit from NIST Standard Reference Database \protect\cite{NIST}.}&29.367 &-  &   &31.774 &-          &           \\
\end{tabular}
\end{ruledtabular}
\end{table*}

%
%

\begin{table*}
\caption{\label{tab:resonances3}Principal quantum numbers $n$, resonance energies (eV), quantum defects
                                                         $\mu$ and autoionization linewidths [$\Gamma$ (meV)], determined from the QB method for
                                                         the dominant $\:3s^23p^4(^1D_2)ns,nd$ and $3s^23p^4(^1S_0)nd$ series resulting from
                                                         photoionization of the $3s^23p^5\,^2P^o_{3/2}$ ground state of Ar$^{+}$.}
\begin{ruledtabular}
\begin{tabular}{cccccccccc}

                    &&      &Rydberg Series         &       &Width ($\Gamma$)   &Rydberg Series     &       & Width ($\Gamma$)\\[.02in]
                    &&      &$\:3s^23p^4(^1D_2)ns,nd$   &       &$\:3s^23p^4(^1D_2)ns,nd$   &$\:3s^23p^4(^1S_0)nd$& & $\:3s^23p^4(^1S_0)nd$\\
Initial State           &&$n$   &Energy (eV)                &$\mu$  & Energy(meV)       &Energy (eV)            &$\mu$  & Energy (meV)  \\[.02in]
\hline
                    &&      &                       &j=5/2  &               &                   &       &           \\
$3s^23p^5\,^2P^o_{3/2}$&&[7s]   &-                      &-      &-              &                   &       &           \\
                    &&8     &27.999                 &1.692  &0.27           &                   &       &           \\
                    &&9     &28.348                 &1.691  &0.12           &                   &       &           \\
                    &&10    &28.579                 &1.690  &0.07           &                   &       &           \\
                    &&11    &28.739                 &1.689  &0.04           &                   &       &           \\
                    &&12    &28.855                 &1.688  &0.03           &                   &       &           \\
                    &&13    &28.941                 &1.688  &0.02           &                   &       &           \\
                    &&14    &29.008                 &1.688  &0.01           &                   &       &           \\
                    &&15    &29.060                 &1.687  &< 10$^{-2}$        &                   &       &           \\
                    &&16    &29.101                 &1.687  &< 10$^{-2}$        &                   &       &           \\
                    &&$\cdot$ &$\cdot$                  &$\cdot$    &               &                   &       &           \\
                    &&$\infty$  &29.367                 &-      &               &                   &       &           \\

                    &&      &                       &j =5/2 &               &                   &       &           \\
$3s^23p^5\,^2P^o_{3/2}$ &&[6d]  &-                      &-      &-              &29.790             &0.735  &25.2       \\
                    &&7     &27.983                 &0.729  &~~2.25         &30.361             &0.749  &14.3       \\
                    &&8     &28.335                 &0.737  &~~1.40         &30.717             &0.757  &~8.8       \\
                    &&9     &28.569                 &0.742  &~~0.94         &30.952             &0.763  &~5.8       \\
                    &&10    &28.731                 &0.745  &~~0.66         &31.116             &0.766  &~4.0       \\
                    &&11    &28.849                 &0.748  &~~0.40         &31.234             &0.769  &~2.9       \\
                    &&12    &28.937                 &0.750  &~~0.33         &31.322             &0.771  &~2.2       \\
                    &&13    &29.004                 &0.751  &~~0.26         &31.390             &0.773  &~1.7       \\
                    &&14    &29.057                 &0.752  &~~0.20         &31.443             &0.774  &~1.3       \\
                    &&15    &29.099                 &0.753  &~~0.16         &31.485             &0.775  &~1.0       \\
                    &&$\cdot$ &$\cdot$                  &$\cdot$    &               &$\cdot$                &$\cdot$    &-          \\

&&$\infty$\footnote{Series limit from NIST Standard Reference Database \protect\cite{NIST}.}&29.367 &-  &   &31.774             &-          &   \\
\end{tabular}
\end{ruledtabular}
\end{table*}

\section{Summary}

Absolute photoionization cross sections have been measured for a
statistical mixture of ground-state and metastable Ar$^+$ in the
photon energy range $\approx$ 27 eV -- 60 eV with a nominal energy
resolution of 10 meV. The cross section in this energy region is
characterized by a wealth of resonance features due to indirect
photoionization. The dominant features were spectroscopically
assigned to Rydberg series and quantum defects were determined. 
A conservative estimate for the accuracy of the resonance energies 
is $\pm$ 10 meV.
The natural linewidths were determined from the R-matrix method 
which we would estimate to be accurate to 10 \%. 
Over the entire energy range investigated, the theoretical results
for the photoionization cross sections on this C$\ell$-like system
obtained from an intermediate coupling Breit-Pauli R-matrix
approach are in excellent agreement with the high resolution
measurements. The main thrust of this joint investigation is the
coupling of high experimental sensitivity and photon energy
resolving power with state-of-the-art theoretical predictions. The
excellent agreement between theory and experiment over the entire
energy range gives added confidence in both results. The
photoionization cross sections from the present study on this
C$\ell$-like ion are suitable for inclusion in state-of-the-art
photoionization modeling codes such as Cloudy
\cite{Ferland1998,Ferland2003} and XSTAR \cite{Kallman2001}, which
are used to numerically simulate the thermal and ionization
structure of ionized astrophysical nebulae.

%
%
%
%
\section{Acknowledgments}

The experimental work was supported by the Office of Basic Energy
Sciences, Chemical Sciences, Geosciences and Energy Biosciences
Division, of the U.S. Department of Energy under grants
DE-FG03-00ER14787 and DE-FG02-03ER15424 with the University of
Nevada, Reno; by the Nevada DOE/EPSCoR Program in Chemical Physics
and by CONACyT through the CCF-UNAM, Cuernavaca, M\'{e}xico.  A. A.
and M. M. S'A. acknowledge support from DGAPA-UNAM-IN 113010 (M\'exico) 
and CNPq (Brazil), respectively. C. P. B. was supported by U.S.
Department of Energy grants  through Auburn
University. B. M. McL. thanks the Institute for Theoretical Atomic and Molecular
Physics (ITAMP) for their hospitality and support under the
visitors program. ITAMP is supported by a grant from the U.S.
National Science Foundation to Harvard University and the
Smithsonian Astrophysical Observatory. The computational work was
carried out at the National Energy Research Scientific Computing
Center in Oakland, CA and on the Tera-grid at the National
Institute for Computational Sciences in Knoxville, Tennessee, which is
supported in part by the U.S. National Science Foundation.
%
%

\bibliographystyle{apsrev}
\bibliography{arplus}

\end{document}